\documentclass{pasj00}
\begin{document}
\SetRunningHead{J.\ Fukue}
{Velocity-Dependent Eddington Factor}
\Received{yyyy/mm/dd}
\Accepted{yyyy/mm/dd}

\title{Velocity-Dependent Eddington Factor
in Relativistic Radiative Flow}

\author{Jun \textsc{Fukue}} 
\affil{Astronomical Institute, Osaka Kyoiku University, 
Asahigaoka, Kashiwara, Osaka 582-8582}
\email{fukue@cc.osaka-kyoiku.ac.jp}



\maketitle


\begin{abstract}
We propose a variable Eddington factor,
depending on the {\it flow velocity} $v$,
for the relativistic radiative flow,
whose velocity becomes of the order of the speed of light.
When the gaseous flow is radiatively 
accelerated up to the relativistic regime,
the velocity gradient becomes very large in the direction of the flow.
As a result, 
the radiative diffusion may become {\it anisotropic}
in the comoving frame of the gas.
Hence, 
in a flow that is accelerated from subrelativistic to relativistic regimes,
the Eddington factor should be different from $1/3$
even in the diffusion limit.
As a simple form,
the velocity-dependent Eddington factor may be written as
$f(\beta) = 1/3+(2/3)\beta$,
where $\beta=v/c$.
Using the velocity-dependent Eddington factor,
we can solve the rigorous equations of the relativistic radiative flow
accelerated up to the relativistic speed.
We also propose a generalized form for a variable Eddington factor
as a function of the optical depth $\tau$ as well as the flow velocity:
$f(\tau, \beta) = 1/3 + (2/3)
[{1+(\tau+1)\beta}]/({1+\tau+\beta})$
for a spherically symmetric case.
The velocity-dependent Eddington factor can be used
in various relativistic radiatively-driven flows,
such as black-hole accretion flows,
relativistic astrophysical jets and outflows, and
relativistic explosions like gamma-ray bursts.
\end{abstract}

\section{Velocity-Dependent Eddington Factor}

In the fundamentals of radiation transfer in the static atmosphere,
in order to solve moment equations truncated at the finite order,
we need some relation to close the sequence
(Chandrasekhar 1960; Mihalas 1970; Rybicki, Lightman 1979;
Mihalas, Mihalas 1984, Shu 1991).
When moment equations are truncated at the second order,
we usually adopt the {\it Eddington approximation}
in the inertial frame (laboratory frame) as a closure relation,
\begin{equation}
   P^{ij} = \frac{\delta^{ij}}{3} E,
\label{PE}
\end{equation}
where
$E$ and $P^{ij}$ are the radiation energy density and
the radiation stress in the inertial frame, respectively.
In a relativistic radiative flow,
where the gas moves at a relativistic speed,
we also assume the Eddington approximation 
in the comoving frame (fluid frame),
\begin{equation}
   P_0^{ij} = \frac{\delta^{ij}}{3} E_0,
\label{P0E0}
\end{equation}
where
$E_0$ and $P_0^{ij}$ are the radiation energy density and
the radiation stress in the comoving frame, respectively.
Then, this closure relation in the comoving frame
is transformed to the relation in the inertial frame
(Lindquist 1966; Hsieh, Spiegel 1976; Fukue et al. 1985;
Kato et al. 1998).

These relations are valid in the diffusion limit,
where the photon mean-free path is sufficiently smaller than
the scalelength of the system in the optically thick regime.
In other words, in such a diffusion limit,
the radiative diffusion is supposed to be {\it isotropic},
and the so-called {\it Eddington factor} $f$ is set to be
\begin{equation}
   f = \frac{1}{3}.
\end{equation}

However, in the optically thin regime,
where the photon mean-free path is sufficiently larger
than the scalelength of the system,
the radiative diffusion generally becomes {\it anisotropic}.
For example, in the spherically symmetric case,
the Eddington factor should be unity in the streaming limit
(it should be noted that in the plane-parallel case
the Eddington factor is always 1/3).
Hence,
in order to bridge the optically thick to thin regimes,
a variable Eddington factor,
which depends on the optical depth $\tau$, is usually adopted,
\begin{equation}
   f(\tau)=\frac{1+\tau}{1+3\tau}
\label{ftau}
\end{equation}
in the spherically symmetric case
(Tamazawa et al. 1975).

Recently, relativistic radiative flows
were rigorously solved under a simple situation by Fukue (2005b).
He examined a flow radiatively-driven
perpendicular to a luminous disk
under a fully special relativistic treatment,
taking into account radiation transfer.
He adopted the current formalism as well as the Eddington approximation
in the comoving frame (e.g., Kato et al. 1998).
When he arranged the basic equations
of relativistic radiation hydrodynamics
for the one-dimensional flow without gravity or other forces,
he found the {\it singularity} in the sense of ``sonic'' points,
where the flow speed is equal to the relativistic sound speed
of $c/\sqrt{3}$.
This singularity originates from the closure relation
(\ref{P0E0}) for the radiation fields,
and the appearance of the singularity suggest that
the Eddington approximation would be invalid
in the relativistic flow with strong velocity gradients
(cf. Turolla, Nobili 1988; Dullemond 1999).

In analogy to the optically thick-thin transition,
the possible invalidity of the Eddington approximation
in such a relativistic flow can be understood as follows.
When the radiative flow is accelerated up to the relativistic regime
and the flow velocity becomes of the order of the speed of light,
the velocity gradient also becomes very large in the direction of flow.
As a result, even in the comoving frame of the gas,
the velocity fields as well as density distributions
are no longer uniform, and
the photon mean-free path would be longer in the downstream direction
than in the upstream or other directions.
Hence, the radiative diffusion becomes {\it anisotropic},
and we should consider a variable Eddington factor:
the Eddington factor may depend on the {\it flow velocity}
$v$ ($=\beta c$) in the form such as
\begin{equation}
   f(\beta) = \frac{1}{3} + \frac{2}{3}\beta.
\label{fbeta}
\end{equation}

In several literatures
(e.g., Mihalas, Mihalas 1984; Mihalas 1986; Kato et al. 1998),
the influence of the flow velocity on radiation transfer
was discussed in the context of the description
between the comoving and inertial frames.
That is, at the lowest order of the flow velocity to the speed of light,
the closure relation in the inertial frame
must be modified to the order of $(v/c)^1$.
In the comoving frame, however,
the Eddington approximation (\ref{P0E0}) was adopted,
and no one considered the possible violation
of the Eddington approximation in the comoving frame.

In this paper,
we thus propose a {\it velocity-dependent variable Eddington factor}
for the relativistic radiative flow,
whose velocity becomes of the order of the speed of light.

In the next section
we describe the basic equations for the one-dimensional
vertical radiative flow,
in order to show the typical problem
which needs the velocity-dependent Eddington factor.
In section 3, we examine several properties
required for the velocity-dependent Eddington factor.
In section 4,
we solve a relativistic radiative flow
under the appropriate boundary conditions
at the flow base and top,
and display the effect of the velocity-dependent Eddington factor.
The final section is devoted to the discussion on
the physical reason of the present problem,
the generalization of the variable Eddington factor,
and concluding remarks.


\section{Basic Equations}

In order to demonstrate the present problem
inherent in the relativistic radiation transfer flow,
we examine a simple one-dimensional flow in what follows.

We consider the plane-parallel case in the vertical direction;
a luminous flat disk or thin skin above a luminous sphere.
The radiative energy is transported in the vertical direction,
and the gas itself also moves in the vertical direction
by the action of radiation pressure.
For simplicity, in the present paper,
the radiation field is sufficiently intense that
both the gravitational field, e.g., of the central object,
and the gas pressure are ignored.
The internal heating is also ignored.
As for the order of the flow velocity $v$,
we consider the fully relativistic regime,
where the terms are retained up to the second order of $(v/c)$.

Under these assumptions,
the radiation hydrodynamic equations
for steady vertical ($z$) flows are described as follows
(Kato et al. 1998; Fukue 2005b).

The continuity equation is
\begin{equation}
   \rho cu = J ~(={\rm const.}),
\label{rho1}
\end{equation}
where $\rho$ is the proper gas density, $u$ the vertical four velocity, 
$J$ the mass-loss rate per unit area,
and $c$ the speed of light.
The four velocity $u$ is related to the proper three velocity $v$ by
$u=\gamma v/c$, where $\gamma$ is the Lorentz factor,
$\gamma=\sqrt{1+u^2}=1/\sqrt{1-(v/c)^2}$.

The equation of motion is
\begin{equation}
   c^2u\frac{du}{dz} = \frac{\kappa_{\rm abs}+\kappa_{\rm sca}}{c}
                    \left[ F \gamma (1+2u^2) - c(E+P)\gamma^2 u \right],
\label{u1}
\end{equation}
where $\kappa_{\rm abs}$ and $\kappa_{\rm sca}$
are the absorption and scattering opacities (gray),
defined in the comoving frame,
$E$ the radiation energy density, $F$ the radiative flux, and
$P$ the radiation pressure observed in the inertial frame.
The first term in the square bracket on the right-hand side
of equation (\ref{u1}) means the radiatively-driven force,
which is modified to the order of $u^2$, whereas
the second term is the radiation drag force,
which is also modified, but roughly proportional to the velocity.

In the no-gas pressure approximation and without heating,
the energy equation is reduced to a radiative equilibrium relation,
\begin{equation}
   0 =  j - c\kappa_{\rm abs} E \gamma^2 - c\kappa_{\rm abs} P u^2
                  + 2 \kappa_{\rm abs} F \gamma u,
\label{j1}
\end{equation}
where $j$ is the emissivity defined in the comoving frame.
In this equation (\ref{j1}),
the third and fourth terms on the right-hand side
appear in the relativistic regime.

For radiation fields, the zeroth-moment equation becomes
\begin{eqnarray}
   \frac{dF}{dz} &=& \rho \gamma
         \left[ j - c\kappa_{\rm abs} E
                 + c\kappa_{\rm sca}(E+P)u^2  \right.
\nonumber
\\
    &&   \left. + c\kappa_{\rm abs}Fu/\gamma
               -\kappa_{\rm sca}F ( 1+v^2/c^2 )\gamma u \right].
\label{F1}
\end{eqnarray}
The first-moment equation is
\begin{eqnarray}
   \frac{dP}{dz} &=& \frac{\rho \gamma}{c} 
         \left[ ju/\gamma - \kappa_{\rm abs} F
                  + c\kappa_{\rm abs}Pu/\gamma \right.
\nonumber
\\
     && \left. -\kappa_{\rm sca}F(1+2u^2)
               +c\kappa_{\rm sca}(E+P)\gamma u \right].
\label{P1}
\end{eqnarray}

In order to close moment equations for radiation fields,
we need some closure relation.
If we adopt the usual Eddington approximation,
\begin{equation}
   P_0 = \frac{1}{3}E_0
\end{equation}
in the comoving frame as the closure relation (Fukue 2005b),
where $P_0$ and $E_0$ are the quantities in the comoving frame,
the transformed closure relation in the inertial frame becomes
\begin{equation}
   cP \left( 1 + \frac{2}{3}u^2 \right) = 
   cE \left( \frac{1}{3} - \frac{2}{3} u^2 \right) 
   + \frac{4}{3} F \gamma u
\label{E1}
\end{equation}
(see Kato et al. 1998 for details).
Instead, we here adopt a {\it velocity-dependent} 
variable Eddington factor
$f(\beta)$,
\begin{equation}
   P_0 = f(\beta) E_0
\label{close0}
\end{equation}
in the comoving frame,
where $\beta=v/c$, $v$ being the flow velocity.
If we adopt this form (\ref{close0}) as the closure relation
in the comoving frame,
the transformed closure relation in the inertial frame is
\begin{equation}
   cP \left( 1 + u^2 - fu^2 \right) = 
   cE \left( f\gamma^2 - u^2 \right) 
   + 2 F \gamma u \left( 1 - f \right),
\label{close}
\end{equation}
or equivalently,
\begin{equation}
   cP \left( 1 - f\beta^2 \right) =
   cE \left( f - \beta^2 \right) + 2F\beta \left( 1 - f \right).
\label{close_beta}
\end{equation}
The constraint on the function $f(\beta)$
will be discussed later.

Eliminating $j$ with the help of equations (\ref{j1}),
and using continuity equation (\ref{rho1}),
equations (\ref{u1}), (\ref{F1}), and (\ref{P1}) are rearranged as
\begin{eqnarray}
\!\!\!\!\!
   cJ\frac{du}{dz} &=& (\kappa_{\rm abs}+\kappa_{\rm sca})
                     \rho \frac{\gamma}{c}
                    \left[ F (1+2u^2) - c(E+P)\gamma u \right],
\label{u2}
\\
\!\!\!\!\!
   \frac{dF}{dz} &=& (\kappa_{\rm abs}+\kappa_{\rm sca})
                   \rho u
                    \left[ c(E+P)\gamma u - F (1+2u^2) \right],
\label{F2}
\\
\!\!\!\!\!
   \frac{dP}{dz} &=& (\kappa_{\rm abs}+\kappa_{\rm sca})
                    \rho \frac{\gamma}{c} 
                    \left[ c(E+P)\gamma u - F (1+2u^2) \right].
\label{P2}
\end{eqnarray}

The integration of the sum of equations (\ref{u2}) and (\ref{P2})
yields the momentum flux conservation along the flow,
\begin{equation}
   cJ u + P = K ~(={\rm const.}).
\label{K}
\end{equation}
Similarly, after some manipulations,
the integration of the sum of equations (\ref{u2}) and (\ref{F2})
gives the energy flux conservation along the flow,
\begin{equation}
   c^2 J \gamma + F = L ~(={\rm const.}).
\label{L}
\end{equation}

At this stage, the basic equations are
the equation of motion (\ref{u2}),
the mass flux (\ref{rho1}), the momentum flux (\ref{K}),
the energy flux (\ref{L}), and the closure relation (\ref{close}).

Next, by introducing the optical depth $\tau$ by
\begin{equation}
    d\tau = - ( \kappa_{\rm abs}+\kappa_{\rm sca} ) \rho dz,
\end{equation}
the equation of motion (\ref{u2}) is rewritten as
\begin{equation}
   cJ\frac{du}{d\tau} = - \frac{\gamma}{c}
                    \left[ F (1+2u^2) - c(E+P)\gamma u \right].
\label{u3}
\end{equation}
Furthermore, eliminating $E$ with the help of equation (\ref{close}),
this equation (\ref{u3}) can be finally rewritten as
\begin{equation}
   cJ\frac{du}{d\tau} = -\frac{\gamma}{c}
           \frac{ F(f\gamma^2+u^2) - cP (1+f) \gamma u}
                {f\gamma^2-u^2},
\label{u}
\end{equation}
or equivalently,
\begin{equation}
   c^2 J \gamma^2 \frac{d\beta}{d\tau} = 
           - \frac{ F(f+\beta^2) - cP (1+f) \beta}{f-\beta^2},
\label{beta}
\end{equation}
where $\beta=v/c$.
When the Eddington factor $f$ is 1/3,
equations (\ref{u}) and (\ref{beta}) are reduced
to those given in Fukue (2005b).

We then solve equations (\ref{u}) [or (\ref{beta})], (\ref{K}), and (\ref{L})
for appropriate boundary conditions at the moving surface,
as discussed in Fukue (2005b), and 
for a suitable form of the variable Eddington factor $f(\beta)$.

\section{Properties of $f(\beta)$}

We here briefly discuss the constraints on
the velocity-dependent variable Eddington factor $f(\beta)$
and choose a preferable candidate for $f(\beta)$.

At first, this function must be reduced to 1/3
in the non-relativistic limit of $\beta=0$.
In the extremely relativistic limit of $\beta=1$
with strong velocity gradients,
where the photon mean-free path becomes large,
this function would approach unity,
as in the case of the streaming limit for an optically thin regime.
Hence, the function $f(\beta)$ must satisfy the boundary conditions,
\begin{eqnarray}
   f(0) &=& \frac{1}{3},
\\
   f(1) &=& 1,
\end{eqnarray}
and should monotonically increase from 0 to 1.

Another constraint comes from the critical conditions.
As can be seen in equations (\ref{u}) and (\ref{beta}),
these equations have {\it singularities} (critical points).
That is, when the condition,
\begin{equation}
    f(\beta)-\beta^2 = 0,
\end{equation}
is satified at some $\beta$,
the denominator of equation (\ref{beta}) vanishes,
and the equation becomes singular.
For example,
if the Eddington factor is fixed as $f=1/3$,
the equation becomes singular at $\beta=\pm 1/\sqrt{3}$,
or at the point, where the flow velocity
is equal to the relativistic sound speed of $c/\sqrt{3}$.
As already pointed out in Dullemond (1999),
this singularity originates from 
the finite number of moments and the closure relation adopted
(see also Fukue 2005b).
In order for the relativistic transfer flow to be always regular,
the condition,
\begin{equation}
    f(\beta)-\beta^2 > 0,
\end{equation}
should be satisfied for $0 \leq \beta < 1$.
Furthermore, since $f=1$ at $\beta=1$,
the equation is marginally critical ($f-\beta^2=0$) at $\beta=1$.

In addition, from the linear analysis around a critical point
of equation (\ref{u}),
the velocity gradient near to the singularity is found to be
\begin{equation}
   \left.
   \frac{du}{d\tau} \right|_{\rm c} =
   \left.
   \frac{u + \frac{\displaystyle F\gamma^3}{\displaystyle c^2J}
             \frac{\displaystyle 1-f}{\displaystyle 1+f}
             \frac{\displaystyle \partial f}{\displaystyle \partial u}}
        {(1-f) 2u - (1+u^2)
             \frac{\displaystyle \partial f}{\displaystyle \partial u}}
             \right|_{\rm c},
\label{dudtau}
\end{equation}
where the subscript c denotes the values at the critical point.
If the Eddington factor is fixed as $f=1/3$,
the velocity gradient becomes $du/d\tau |_{\rm c}=3/4$
($d\beta/d\tau |_{\rm c}=1/\sqrt{3}$).
Or, the transonic solution for the outflow ($u>0$) 
must be decelerated at the critical point,
as was discussed in Fukue (2005b),
where the subsonic solutions were obtained.
Hence, in order for the flow to be accelerated,
even if the solution is marginally critical ($f-\beta^2=0$) at $\beta=1$,
the velocity gradient $du/d\tau |_{\rm c}$ 
should be negative (or zero) there
for the outflow ($u>0$) solutions.
Since the numerator on the right-hand side of equation (\ref{dudtau})
is always positive,
the velocity gradient is negative if the denominator
of equation (\ref{dudtau}) is negative:
\begin{equation}
   \left. \frac{\partial}{\partial u}
   \left[ (1-f)(1+u^2) \right] \right|_{\rm c}
   = \left. \frac{\partial}{\partial u}
   \left[ (1-f)\gamma^2 \right] \right|_{\rm c} <0,
\end{equation}
or equivalently,
\begin{equation}
   \frac{\partial}{\partial \beta}
   \left. \left( \frac{1-f}{1-\beta^2} \right) \right|_{\rm c} <0.
\end{equation}

Now, bearing these constraints in mind,
we examine several possible candidates
of velocity-dependent Eddington factors.

One simple case is a form of
\begin{equation}
   f(\beta) = \frac{1}{3}+\frac{2}{3}\beta^n,
\end{equation}
where $n$ is constant.
In this case, the Eddington factor is a monotonically increasing
function of $\beta$, 
which satifies the boundary conditions at $\beta=0$ and 1.
However, since $f-\beta^2$ becomes negative at some $\beta$ 
for large values of $n$ ($n \geq 3$),
the values of $n$ should be restricted as $n \leq 3$.
Moreover, for this form,
the equation is marginally critical at $\beta=1$,
and the velocity gradient at the critical point is
negative, zero, and positive for $n=1$, 2, 3, respectively.
Hence, the case of $n=3$ is rejected,
and the cases of $n=1$ and 2 are retained;
the simplest case is $n=1$.

Alternative is a form of
\begin{equation}
   f(\beta) = \frac{1}{3-2\beta^n},
\end{equation}
where $n$ is constant.
In this case, $f$ is also monotonically increasing,
but $f-\beta^2$ becomes negative for large $n$ ($n \geq 2$),
and the values of $n$ may be around unity.
Moreover, for this form,
the equation is marginally critical at $\beta=1$,
and the velocity gradient at the critical point is positive,
and this form may be inadequate.

In analogy to a variable Eddington factor
for the optically thick-thin transition,
we can consider a form of
\begin{equation}
    f(\beta) = \frac{1+1/u}{1+3/u}
             = \frac{\beta+\sqrt{1-\beta^2}}{\beta+3\sqrt{1-\beta^2}}.
\end{equation}
In this case, $f$ is monotonically increasing,
but $f-\beta^2$ becomes negative at large $u$.
Similar forms such as
$f=(1+1/u^n)/(1+3/u^n)$ or
$f=[1+1/(\gamma-1)^n]/[1+3/(\gamma-1)^n]$
would be also inadequate,
although we have not checked all possible cases one by one.

Consequently,
we adopt the simplest case of
\begin{equation}
   f(\beta) = \frac{1}{3} + \frac{2}{3} \beta
\label{Eddington}
\end{equation}
in this paper.

\section{Examples of Solutions}

In this section
we show several examples of solutions
of equations (\ref{beta}), (\ref{K}), and (\ref{L})
with velocity-dependent Eddington factors
under appropriate boundary conditions at the moving surface.

As already pointed out in Fukue (2005b),
the usual boundary conditions for the static atmosphere
cannot be used for the present radiative flow,
which moves with velocity at the order of the speed of light.

When there is no motion in the gas (``static photosphere''),
the radiation field just above the surface
under the plane-parallel approximation is easily obtained.
Namely, just above the disk with surface intensity $I_{\rm s}$,
the radiation energy density $E_{\rm s}$, 
the radiative flux $F_{\rm s}$, and
the radiation pressure $P_{\rm s}$ are
$(2/c)\pi I_{\rm s}$,
$\pi I_{\rm s}$, and 
$(2/3c)\pi I_{\rm s}$, respectively,
where the subscript s denotes the values at the disk surface.
However,
the radiation field just above the surface changes
when the gas itself does move upward (``moving photosphere''),
since the direciton and intensity of radiation
change due to relativistic aberration and Doppler effect
(cf. Kato et al. 1998; Fukue 2000).

Let us suppose a situation that
a flat infinite photosphere with surface intensity $I_{\rm s}$
in the comoving frame is not static,
but moving upward with a speed $v_{\rm s}$ 
($=c\beta_{\rm s}$, and
the corresponding Lorentz factor is $\gamma_{\rm s}$),
where the subscript s denotes the values at the surface.
Then, just above the surface,
the radiation energy density $E_{\rm s}$, 
the radiative flux $F_{\rm s}$, and
the radiation pressure $P_{\rm s}$ measured in the inertial frame
become, respectively,
\begin{eqnarray}
   cE_{\rm s} 
   &=& {2\pi I_{\rm s}}
       \frac{3\gamma_{\rm s}^2+3\gamma_{\rm s}u_{\rm s}+u_{\rm s}^2}{3},
\label{Es2}
\\
   F_{\rm s}
   &=& {2\pi I_{\rm s}}
       \frac{3\gamma_{\rm s}^2+8\gamma_{\rm s}u_{\rm s}+3u_{\rm s}^2}{6},
\label{Fs2}
\\
   cP_{\rm s}
   &=& {2\pi I_{\rm s}}
       \frac{\gamma_{\rm s}^2+3\gamma_{\rm s}u_{\rm s}+3u_{\rm s}^2}{3},
\label{Ps2}
\end{eqnarray}
where $u_{\rm s}$ ($=\gamma_{\rm s}v_{\rm s}/c$)
is the flow four velocity at the surface (Fukue 2005b).

Thus, we impose the following boundary conditions:
At the flow base (deep ``inside'' the atmosphere)
with an arbitrary optical depth $\tau_0$,
the flow velocity $u$ is zero,
the radiative flux is $F_0$
(which is a measure of the strength of radiation field), and
the radiation pressure is $P_0$
(which connects with the radiation pressure gradient
and relates to the internal structure),
where the subscript 0 denotes the values at the flow base.
At the flow top (``surface'' of the atmosphere)
where the optical depth is zero,
the radiation field should satisfy the values
above a moving photosphere given by
equations (\ref{Es2})--(\ref{Ps2}).

Applying boundary conditions (\ref{Es2})--(\ref{Ps2})
to equations (\ref{K}) and (\ref{L}),
we have two relations on the boundary values and mass-loss rate:
\begin{eqnarray}
   Jc^2 u_{\rm s} + cP_{\rm s} &=& cP_0,
\label{bc1}
\\
   Jc^2 \gamma_{\rm s} + F_{\rm s} &=& Jc^2 + F_0.
\label{bc2}
\end{eqnarray}

Physically speaking,
in the radiative flow starting from the flow base
with an arbitrary optical depth $\tau_0$,
for initial values of $F_0$ and $P_0$ at the flow base,
the final values of
the radiation fields $E_{\rm s}$, $F_{\rm s}$, $P_{\rm s}$, and
the flow velocity $v_{\rm s}$ at the flow top
can be obtained by solving basic equations.
Furthermore, the mass-loss rate $J$
is determined as an eigenvalue
so as to satisfy the bondary condition at the flow top
(cf. Fukue 2005a in the subrelativistic regime).

In the present, fully relativistic case, however,
the final values of the radiation fields at the flow top depend
on the flow velocity there,
and the final values at the flow top
cannot be analytically expressed by the initial values at the flow base.
Hence, in this paper
we determine the mass-loss rate as follows.

\begin{figure}
  \begin{center}
  \FigureFile(80mm,80mm){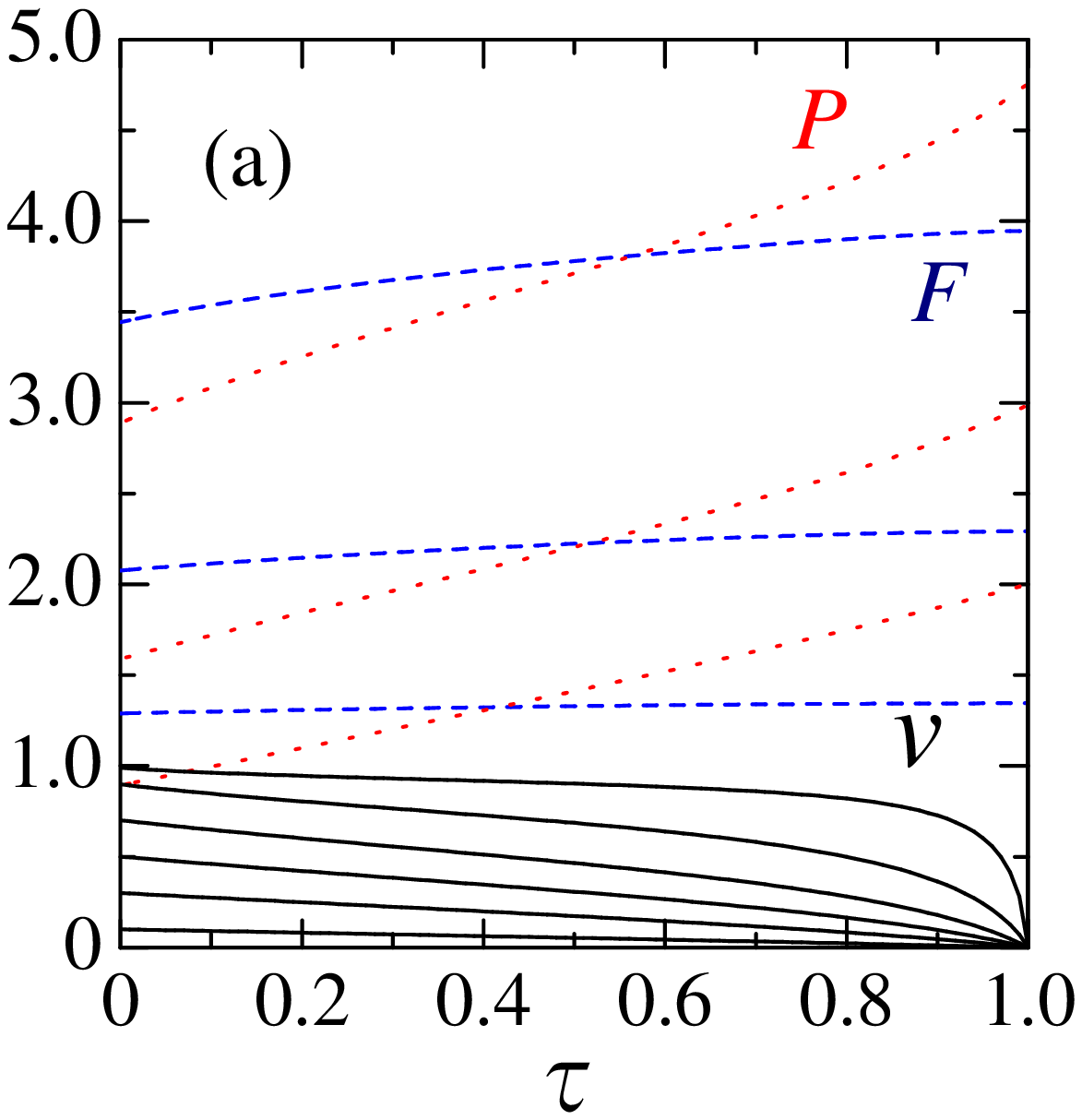}
  \end{center}
  \begin{center}
  \FigureFile(80mm,80mm){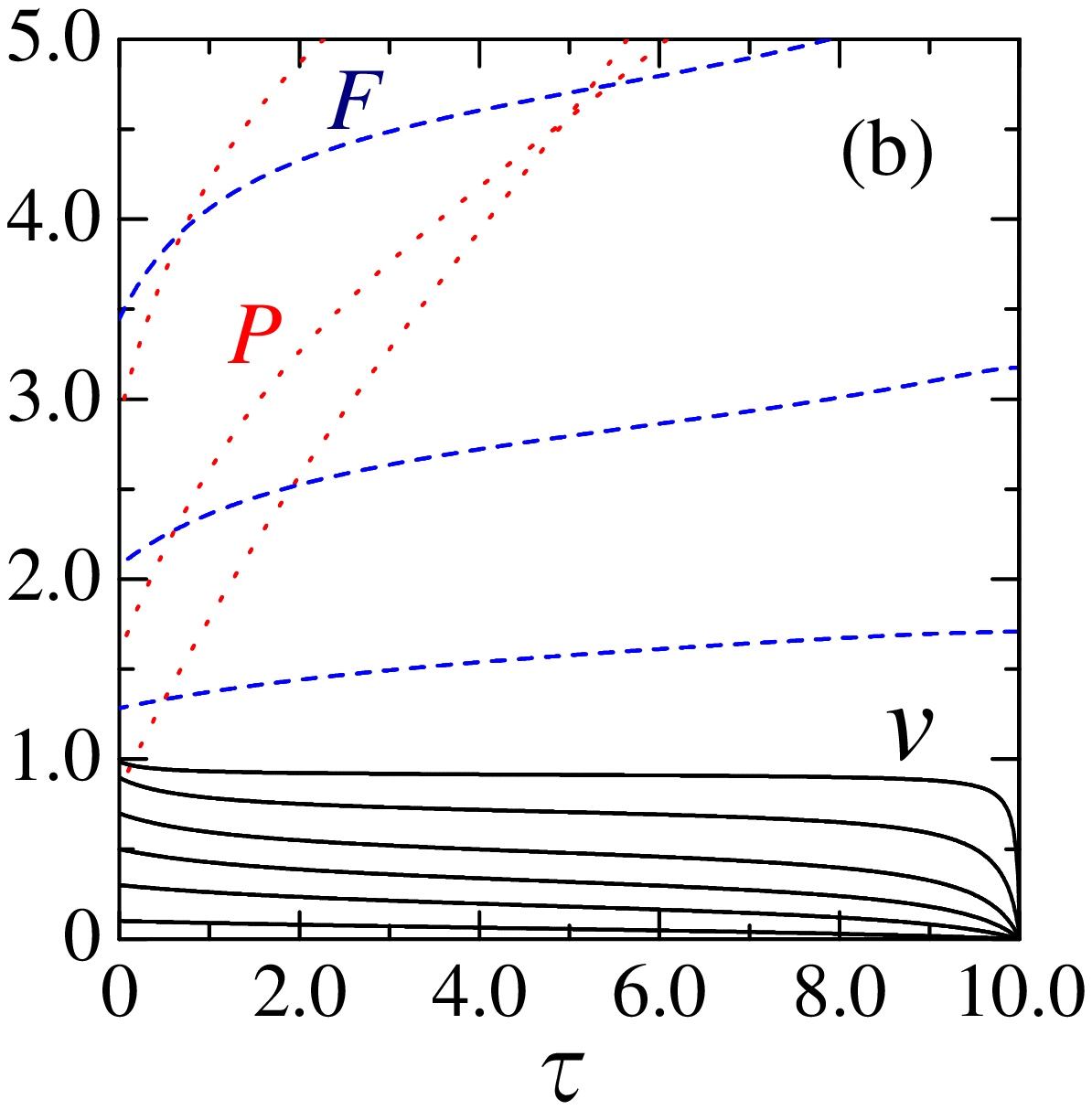}
  \end{center}
\caption{
Flow three velocity $v$ (solid curves), 
radiative flux $F$ (dashed curves),
and radiation pressure $P$ (doted curves),
as a function of the optical depth $\tau$
for several values of $v_{\rm s}$ at the flow top
in a few cases of $\tau_0$:
(a) $\tau_0=1$ and (b) $\tau_0=10$.
The values of $v_{\rm s}$ are
0.1, 0.3, 0.5, 0.7, 0.9, and 0.99
 from bottom to top of $v$
and 0.1, 0.3, and 0.5 from bottom to top of $F$ and $P$.
The velocity $v$ is in units of $c$,
$F$ and $cP$ in $\pi I_{\rm s}$.
}
\end{figure}

In the radiative flow with optical depth $\tau_0$,
we first give the final flow velocity $v_{\rm s}$ (and $\gamma_{\rm s}$),
instead of the initial value of $P_0$.
Then, the final values of radiation fields $E_{\rm s}$, $F_{\rm s}$,
and $P_{\rm s}$ can be fixed by equations (\ref{Es2})--(\ref{Ps2}).
Next, we give a trial value for the mass-loss rate $J$,
and the initial values of $P_0$ and $F_0$
can be fixed by equations (\ref{bc1}) and (\ref{bc2}).
Since all the parameters are temporarily fixed,
we solve equation (\ref{beta}) from $\tau=\tau_0$ to $\tau=0$.
Generally, however, the obtained final velocity at $\tau=0$
is different from a given $v_{\rm s}$.
Thus, we vary the value of $J$
and follow iterative processes,
so that the calculated final velocity coincides 
with a given final velocity $v_{\rm s}$.

Finally, we adopt the velocity-dependent Eddington factor
of the form (\ref{Eddington}).

\begin{figure}
  \begin{center}
  \FigureFile(80mm,80mm){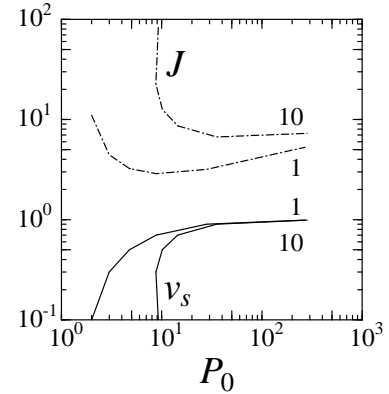}
  \end{center}
\caption{
Final velocity $v_{\rm s}$ at the flow top (solid curves),
and the mass-loss rate $J$ (chain-dotted ones),
as a function of $P_0$
for several values of $\tau_0$ at the flow base:
$\tau_0=1$ and 10.
The quantities are normalized in units of $c$ and $\pi I_{\rm s}$.
That is, the unit of $F$ and $cP$ is $\pi I_{\rm s}$
and the unit of $J$ is $\pi I_{\rm s}/c^2$.
}
\end{figure}

Examples of the relativistic radiative flows
under the present velocity-dependent Eddington factor 
are shown in figures 1 and 2.

In figure 1
we show the flow three velocity $v$ (solid curves),
the radiative flux $F$ (dashed curves),
and the radiation pressure $P$ (doted curves),
as a function of the optical depth $\tau$
for several values of $v_{\rm s}$ at the flow top
and for $\tau_0=1$ and 10.
The quantities are normalized in units of $c$ and $\pi I_{\rm s}$.

In contrast to Fukue (2005b),
where the acceleration solution beyond $c/\sqrt{3}$
was not obtained by the presence of the singularity,
originating from the usual fixed Eddington factor,
the present flow velocity can easily exceed $c/\sqrt{3}$
to be accelerated toward the highly relativistic regime,
since the singularity is evaded
by the adoption of the velocity-dependent Eddington factor (\ref{Eddington}).

Other properties of the flow are similar
to the previous case.
For example,
when the initial radiative flux $F_0$ at the flow base is large,
the flow is effectively accelerated, and
the final speed of the flow becomes large.
The mass-loss rate $J$ is not given arbitrarily,
but is determined as an eigenvalue (Fukue 2005a, b).

In figure 2
we show the final velocity $v_{\rm s}$ at the flow top (solid curves)
and the mass-loss rate $J$ (chain-dotted ones),
as a function of $P_0$
for several values of $\tau_0$ at the flow base.
The quantities are normalized in units of $c$ and $\pi I_{\rm s}$.
For example, the unit of $J$ is $\pi I_{\rm s}/c^2$.

As can be seen in figure 1,
as the radiative flux increases,
the final flow velocity at the flow top increases.
Moreover,
as can be seen in figure 2, 
and similar to the previous case (Fukue 2005a, b),
in order for the flow to exist,
the radiation pressure $P_0$ at the flow base is restricted
in some range.
In the subrelativistic case without gravity and heating (Fukue 2005a),
the initial pressure $P_0$ is proved to be restricted in the range of
$2/3 < {cP_0}/{F_{\rm s}} < 2/3 +  \tau_0$.
In the present case,
the initial pressure is also restricted in some range,
but is modified due to the relativistic effect
and the velocity-dependent Eddington factor.
In addition,
the loaded mass increases as the initial optical depth increases,
even if the final flow speeds are the same.

\begin{figure}
  \begin{center}
  \FigureFile(80mm,80mm){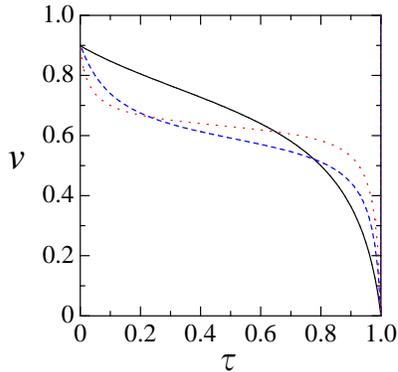}
  \end{center}
\caption{
Flow three velocity $v$ 
in several forms of velocity-dependent Eddington factors.
The parameters are
$\tau_0=1$ and $v_{\rm s}=0.9~c$.
}
\end{figure}

For comparison, 
we show in figure 3 the results for several other forms
of velocity-dependent Eddington factors.
In figure 3,
the parameters are fixed as
$\tau_0=1$ and $v_{\rm s}=0.9~c$.
A solid curve is for the case of $f(\beta)=1/3+2\beta/3$,
whereas
a dashed one is for the case of $f(\beta)=1/3+2\beta^2/3$,
and
a dotted one is for the case of $f(\beta)=1/(3-2\beta)$.
The boundary conditions at the flow top are the same in each case,
but the radiation fields ($F_0$ and $P_0$) at the flow base
and the loaded mass $J$ are different in each case.
In particular,
the loaded mass in the case of $f(\beta)=1/3+2\beta/3$
is larger than those in other cases examined,
and this is another reason to prefer the present form.

It should be finally noted that,
thanks to the velocity-dependent Eddington factor,
the relativistic radiative flow
can be accelerated to reach an extremely relativistic regime,
and the barrier of the magic speed above the static atmosphere
(Icke 1989) would be now completely cleared away.

\section{Discussion and Concluding Remarks}

Moment equations for relativistic radiation transfer
have been derived in several literature
(Lindquist 1966; Anderson, Spiegel 1972; Hsieh, Spiegel 1976;
Thorne 1981; Udey, Israel 1982; Schweizer 1982).
A complete set of moment equations for relativistic flow
is given by the projected, symmetric, trace-free formalism
by Thorne (1981).
Since a moment expansion gives an infinite set of equations,
one must truncate the expansion at the finite order
by adopting a suitable closure assumption,
in order to make the transfer problem tractable.
When one truncate at the second order, for example,
the Eddington approximation is usually adopted
as a closure relation.

Such radiation moment formalism is quite convenient,
and it is a powerful tool for tackling problems
of relativistic radiation hydrodynamics.
However, its validity is never known
unless fully angle-dependent radiation transfer equation
is solved.

Actually, the pathological behavior
in relativistic radiation moment equations
has been pointed out
(Turolla, Nobili 1988; Dullemond 1999).
Namely, moment equations for relativistic radiation transfer
can have singular (critical) points.
For example, in the one-dimensional relativistic radiation flow
using the Eddington approximation (\ref{P0E0}),
where the moment equations are truncated at the second order,
the singularity appears when the flow velocity becomes
$\beta=v/c=1/\sqrt{3}$ (e.g., Fukue 2005b).
As a result, solutions behave pathologically
in regions of strong velocity gradients.
The appearance of singularities is explained
because we approximate the full transfer equations
with a finite number of moments (Dullemond 1999).

In order to avoid the singularity, in this paper,
we thus propose and examine a {\it velocity-dependent
variable Eddington factor}
in radiation transfer in relativistic radiative flows,
which are accelerated up to a relativistic speed
with strong velocity gradients.

As already stated in section 1,
the physical reason is clear.
In an optically thick, low velocity regime,
the mean free path of photons are the same in all directions,
and the radiative diffusion is isotropic.
In a relativistically accelerated flow, however,
the velocity gradient of the flow becomes very large,
and the density distribution also becomes non-uniform
even in the comoving frame of the gas.
Hence,
the mean free path becomes longer in the downstream direction
than in the upstream and other directions,
even in an optically thick regime,
and the radiative diffusion becomes anisotropic.
Thus, in the case of subrelativistic to relativistic regimes,
as in the case of optically thick to thin regimes,
we should consider a variable Eddington factor
which depends on the flow velocity $v$.
The most preferable one, we propose, is
\begin{equation}
   f(\beta)=\frac{1}{3} + \frac{2}{3}\beta,
\label{fbeta41}
\end{equation}
where $\beta=v/c$.

By adopting the velocity-dependent Eddington factor,
the singularity found under the traditional formalism is removed,
and the relativistic radiative flow
can be solved from a low velocity regime
to an extremely relativistic regime.
It should be noted, however, that
the present approach is relevant only for moment expansion
up to the second order.
If we attempt to solve the moment equation up to higher order,
we may search alternative ways.

In this paper, we considered only the one-dimensinal case.
The vector form of the present variable Eddington factor
may be written as
\begin{equation}
   \mbox{\boldmath $f$}(\mbox{\boldmath $\beta$}) = \left(
             \frac{1}{3}-\frac{1}{3}\beta,
             \frac{1}{3}-\frac{1}{3}\beta,
             \frac{1}{3}+\frac{2}{3}\beta \right),
\end{equation}
when the velocity components are 
$(0, 0, \beta)$.

We also propose a general form of a variable Eddington factor
$f(\tau, \beta)$,
which would be useful for an arbitrary optical depth $\tau$,
and for an arbitrary velocity $v$ ($=\beta c$).
The conditions for such a generalized Eddington factor are:
(i) in the low velocity limit ($\beta \sim 0$),
it has the similar form as equation (\ref{ftau}),
(ii) in the optically thick limit ($\tau \gg 1$),
it becomes the presently proposed form,
(iii) in the optically thick, low velocity limit,
it is 1/3,
and
(iv) in the optically thin limit ($\tau \sim 0$) and/or
in the relativistic limit ($\beta \sim 1$),
it approaches unity.
For a spherically symmetric case,
in order to satisfy these conditions,
we apply the relativistic summation rule for 
the flow speed $v$ and
the photon diffusion speed $c/(\tau+1)$, and obtain
a generalized Eddington factor as
\begin{equation}
   f(\tau, \beta) =
      \frac{1}{3}+\frac{2}{3}
         \frac{ \frac{\displaystyle 1}{\displaystyle \tau+1} + \beta}
              {1 + \frac{\displaystyle \beta}{\displaystyle \tau+1}}
   =     \frac{1}{3}+\frac{2}{3}
         \frac{1+(\tau+1)\beta}{1+\tau+\beta}.
\label{ftaubeta}
\end{equation}
It should be noted that, in the low velocity limit ($\beta \sim 0$),
this hybrid form (\ref{ftaubeta}) becomes
$f(\tau, 0) = (3+\tau)/(3+3\tau)$,
which is somewhat different with equation (\ref{ftau}),
but essentially the same.
In the optically thick limit ($\tau \gg 1$),
this equation (\ref{ftaubeta}) is reduced to equation (\ref{Eddington}).
For a plane-parallel case, on the other hand,
equation (\ref{fbeta41}) can be used
for an arbitrary optical depth.

In the present paper,
we examine the plane-parallel case
with the form (\ref{fbeta41}).
The spherically symmetric case
using the hybrid form (\ref{ftaubeta})
will be discussed in a seperate paper.

The velocity-dependent variable Eddington factor
for relativistic radiative flows proposed in the present paper
is fundamentally important
in various aspects of relativistic astrophysics with radiation transfer.
For example,
the present form may be applied to the cases of
black-hole accretion flows with supercritical accretion rates,
relativistic jets and winds driven by luminous central objects,
relativistic explosions including gamma-ray bursts,
neutrino transfers in supernova explosions,
and various events occured in the proto universe.

Of course, the present proposition is only the first one,
and not the final one.
There exist many points to be improved,
including more suitable forms for velocity-dependent Eddington factors.
For example, the Eddington factor
may depend not on the velocity, but on the velocity gradient.
In this field of relativistic radiation tranfer,
there still remain many problems to be solved.

\vspace*{1pc}

The authors would like to thank Professor S. Kato
for valuable comments and discussions.
This work has been supported in part
by a Grant-in-Aid for Scientific Research (15540235 JF) 
of the Ministry of Education, Culture, Sports, Science and Technology.


\end{document}